\documentclass{article}
\usepackage[a4paper, margin=2.5cm]{geometry}
\usepackage{graphicx} % Required for inserting images
\usepackage[dvipsnames]{xcolor}
\usepackage{hyperref}
\usepackage[labelfont=bf]{caption}
\usepackage{booktabs}
\usepackage{amsmath, amsfonts, amssymb, textcomp}
\usepackage{float}
\usepackage{comment}
\usepackage{siunitx}
\usepackage{makecell}
\usepackage{multirow}
\usepackage{pdflscape}
\usepackage{tabularx}
\usepackage{array}
\usepackage[table]{xcolor}
\usepackage{authblk}
\usepackage{upgreek}

\newcolumntype{Y}{>{\centering\arraybackslash}X}

\newcounter{suppfigure}
\title{A wearable, stretchable radio-frequency coil for extremity imaging in low-field MRI systems}
\author[1]{Jesús Conejero}
\author[1]{Marina Fernández-García}
\author[1]{Jose Borreguero}
\author[1,2]{Teresa Guallart-Naval}
\author[1]{Pablo García-Cristóbal}
\author[1]{Lorena Vega-Cid}
\author[1]{Eli G. Castanon}
\author[1]{Rubén Bosch}
\author[1]{Eduardo Pallás}
\author[3]{Pablo Moreno}
\author[1]{Fernando Galve}
\author[1]{Joseba Alonso}
\author[1,3]{José Miguel Algarín}

\affil[1]{The Institute for Instrumentation in Molecular Imaging (I3M), CSIC–Universitat Politècnica de València, Valencia, Spain}
\affil[2]{C.J. Gorter MRI Center, Department of Radiology, Leiden University Medical Center, Leiden, Netherlands}
\affil[3]{PhysioMRI Tech, Paterna, Spain}
\date{September 2026}

\begin{document}

\maketitle

\section*{Abstract}

\textbf{Purpose:} To develop a wearable, stretchable radio-frequency (RF) coil that conforms to the extremities and improves the filling factor in low-field MRI.

\textbf{Methods:} A stretchable solenoid RF coil was fabricated by stitching a sinusoidally arranged Litz-wire conductor onto an elastic textile. The coil was compared with four rigid coils, including two routinely used designs and two prototypes designed to isolate the effects of conductor material and stretchability. Performance was characterized through quality-factor, loading-factor, transmit-efficiency, and image-based signal-to-noise ratio (SNR) measurements. Phantom and \emph{in-vivo} knee experiments for one volunteer were performed using two portable MRI systems operating at 3.53 and 3.04~MHz.

\textbf{Results:} The electrical performance of the selected Litz wire decreased with increasing frequency, becoming a slight disadvantage at 3.53\,MHz. Nevertheless, the stretchable coil provided the highest phantom SNR in both systems, exceeding that of the best-performing rigid coil by approximately 8\,\% at 3.53\,MHz and 20\,\% at 3.04\,MHz. \emph{In vivo}, its global SNR was 5\,\% lower than that of the best rigid coil at 3.53\,MHz and 19\,\% higher at 3.04\,MHz. Compared with the larger rigid coil required when knee positioning is constrained, the corresponding SNR improvements were approximately 18\,\% and 80\,\%.

\textbf{Conclusion:} Anatomical conformity can compensate for the frequency-dependent electrical limitations of Litz wire in stretchable low-field RF coils. The proposed design provided SNR comparable to or greater than the best rigid designs while facilitating coil placement in subjects for whom smaller rigid coils may be impractical.

\section{Introduction}

Radio-frequency (RF) coils play a critical role in determining the signal-to-noise ratio (SNR) and overall imaging performance of low-field magnetic resonance imaging (MRI) systems \cite{hoult1976signal,gruber2018rf}. The recent resurgence of low-field MRI has created new opportunities for affordable, portable, and point-of-care imaging. However, the reduced nuclear magnetization at low field strengths results in inherently weaker signals than in conventional high-field systems \cite{guallart2022portable,arnold2023low,wald2020low,geethanath2019accessible,sarracanie2015low}. SNR is therefore a primary constraint on the achievable image quality. Maximizing RF coil sensitivity and coupling the coil efficiently to the anatomy are essential for obtaining clinically useful images. In particular, coils that conform closely to the patient can increase the filling factor and improve sensitivity \cite{gruber2018rf}.

Solenoids are widely used as volume transmit-receive (Tx/Rx) coils in low-field MRI because they can generate a roughly homogeneous $B_1$ field perpendicular to the static $B_0$ field \cite{gruber2018rf,webb2023tackling,coffey2013low,giovannetti2022radiofrequency}. This configuration is particularly suitable for permanent-magnet systems based on Halbach arrays \cite{o2021vivo,galve2024elliptical}. Other geometries, including saddle and toroidal coils, have been developed for specific field orientations or to reduce sensitivity to external electromagnetic interference \cite{bidinosti2005active,vliem2026design}. Nevertheless, RF coils for low-field MRI are generally implemented as rigid structures.

Rigid extremity coils present an inherent compromise between patient access and sensitivity. For knee or wrist imaging, the coil must be sufficiently large to allow the foot or hand to pass through its interior. Its diameter may therefore substantially exceed that of the target anatomy, reducing the filling factor and limiting SNR \cite{hoult1976signal,gruber2018rf}. This limitation is particularly relevant for patients with restricted mobility and for coils intended to accommodate a wide range of anatomical sizes.

Flexible and stretchable coils that conform to the region of interest are now commonplace in high-field MRI \cite{vincent2019stitching,port2020detector}. These designs have been classified as adaptable, wearable, or stretchable according to their mechanical implementation \cite{ramesh2025adaptable}. By reducing the separation between the conductor and the anatomy, conformable coils can increase the filling factor and improve sensitivity \cite{hoult1976signal,gruber2018rf}. This approach may be especially valuable in low-field MRI, where SNR is limited, but its benefits remain insufficiently characterized.

One possible conductor for wearable low-field RF coils is Litz wire, which comprises multiple individually insulated strands twisted or woven into a single conductor \cite{webb2023tackling,mispelter2006nmr,zhang2014realistic,geng2021modelling,guillod2017litz}. Its mechanical flexibility facilitates integration into conformable structures \cite{stormont2023flexible}. Electrically, strand insulation and transposition can reduce losses associated with skin and proximity effects, thereby decreasing the AC resistance and potentially improving the coil quality factor and efficiency \cite{webb2023tackling,coffey2013low,mispelter2006nmr,zhang2014realistic,geng2021modelling,guillod2017litz,sullivan2014simplified,giovannetti2023conductor,lotfi1993high}.

The electrical performance of Litz wire is strongly frequency- and construction-dependent. As the operating frequency increases and the skin depth decreases, maintaining low AC resistance requires increasingly thin, well-transposed strands. Such strands may be more susceptible to mechanical damage during repeated deformation. Litz-wire selection therefore involves a compromise among strand geometry, electrical performance, flexibility, and mechanical robustness \cite{mispelter2006nmr,giovannetti2017litz,pauly2006optimized}. Previous low-field MRI implementations include a form-fitting head coil with lower AC resistance than an equivalent solid-copper coil \cite{sarracanie2015low}, a flexible wearable quadrature knee coil \cite{wan2023flexible}, and rigid solenoids in which Litz wire improved $Q$ and phantom SNR relative to solid copper conductors \cite{webb2023tackling}. However, the respective contributions of the conductor and the conformable mechanical implementation have not been systematically separated.

In this work, we develop and evaluate a wearable, stretchable solenoid RF coil for extremity imaging in portable low-field MRI. The coil combines a sinusoidally arranged Litz-wire conductor with an elastic textile support, allowing it to be placed around the limb as a garment. We compare it with four rigid coils designed to distinguish the effects of conductor material and stretchability. The coils are characterized through quality-factor, loading-factor, transmit-efficiency, and imaging SNR measurements in two portable MRI systems operating at 3.53\,MHz and 3.04\,MHz. This study examines whether the increased anatomical conformity of a stretchable design can compensate for any electrical penalties associated with its conductor and mechanical implementation.

%%%%%%%%%%%%%%%%%%%%%%%%%%%%%%%%%%%%% METHODS %%%%%%%%%%%%%%%%%%%%%%%%%%%%%%%%%%%%%

\section{Materials \& Methods}

\subsection{Scanners}

\begin{figure}
    \centering
    \includegraphics[width=0.5\linewidth]{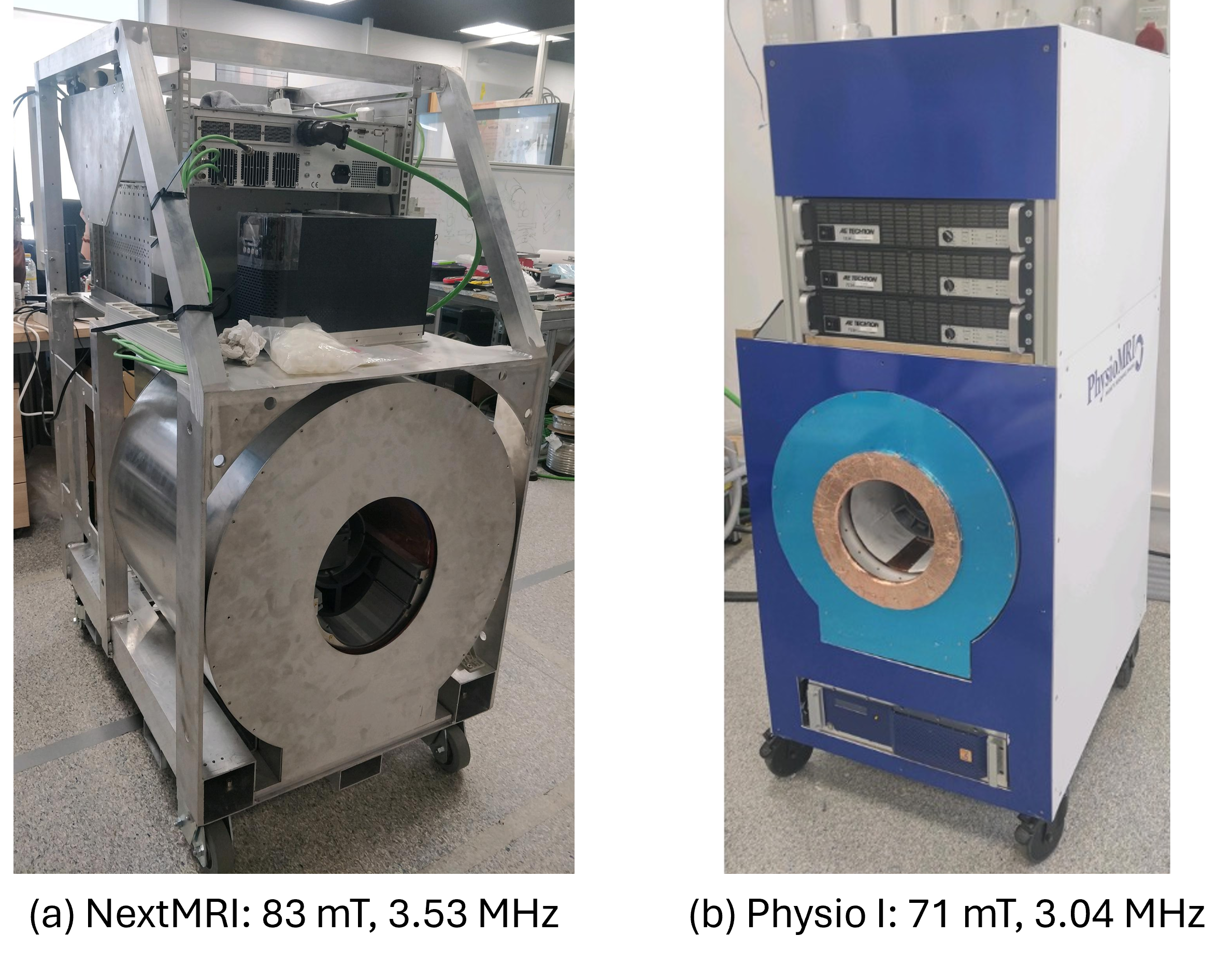}
    \caption{Portable Halbach MRI scanners used in this work. (a) NextMRI, operating at 83\,mT and 3.53\,MHz. (b) Physio I, operating at 71\,mT and 3.04\,MHz.}
    \label{fig:scanners}
\end{figure}

Experiments were performed using two portable MRI scanners. The first, hereafter referred to as NextMRI, operates at approximately 83\,mT, corresponding to a $^1$H Larmor frequency of approximately 3.53\,MHz. The scanner uses an elliptical Halbach permanent magnet as the main magnetic field source \cite{galve2024elliptical} (Figure \ref{fig:scanners}a). The second scanner operates at approximately 71\,mT, corresponding to a $^1$H Larmor frequency of approximately 3.04\,MHz, and uses a circular Halbach array \cite{guallart2022portable}. This scanner is hereafter referred to as Physio I (Figure \ref{fig:scanners}b).

Both scanners routinely use rigid solenoid RF coils with diameters of 15\,cm and 18\,cm, wound using solid copper wire. The scanners are operated using the open-source Magnetic Resonance Control System (MaRCoS) \cite{NEGNEVITSKY2023107424,guallart2023benchmarking}, with user interaction provided through the MaRCoS Graphical Environment (MaRGE) \cite{algarin2024marge}.

\subsection{Coil fabrication}

\subsubsection{Selection of the Litz wire}

The Litz wire was selected by considering both its electrical performance and mechanical characteristics. To exploit the electrical advantages of Litz wire, the strand diameter should be smaller than or comparable to the skin depth, while the conductor should contain a sufficiently large number of appropriately transposed strands \cite{webb2023tackling,guillod2017litz,stormont2023flexible,sullivan2014simplified,giovannetti2023conductor,giovannetti2017litz}. At the operating frequencies of the two scanners, the skin depth in copper is approximately 35\,$\upmu$m. Mechanically, the conductor must be sufficiently flexible to follow the required path and accommodate repeated deformation of the stretchable coil.

Based on these electrical and mechanical considerations, we selected a Litz wire comprising 1,500 strands, each with a diameter $d_\text{strand}=30\,\upmu$m, manufactured by Elektrisola (Reichshof-Eckenhagen, Germany). This construction provided a suitable compromise between the strand-diameter criterion at the operating frequencies and the flexibility required for the wearable stretchable design.

\subsubsection{Fabrication procedure}

\begin{figure}
    \centering
    \includegraphics[width=1\linewidth]{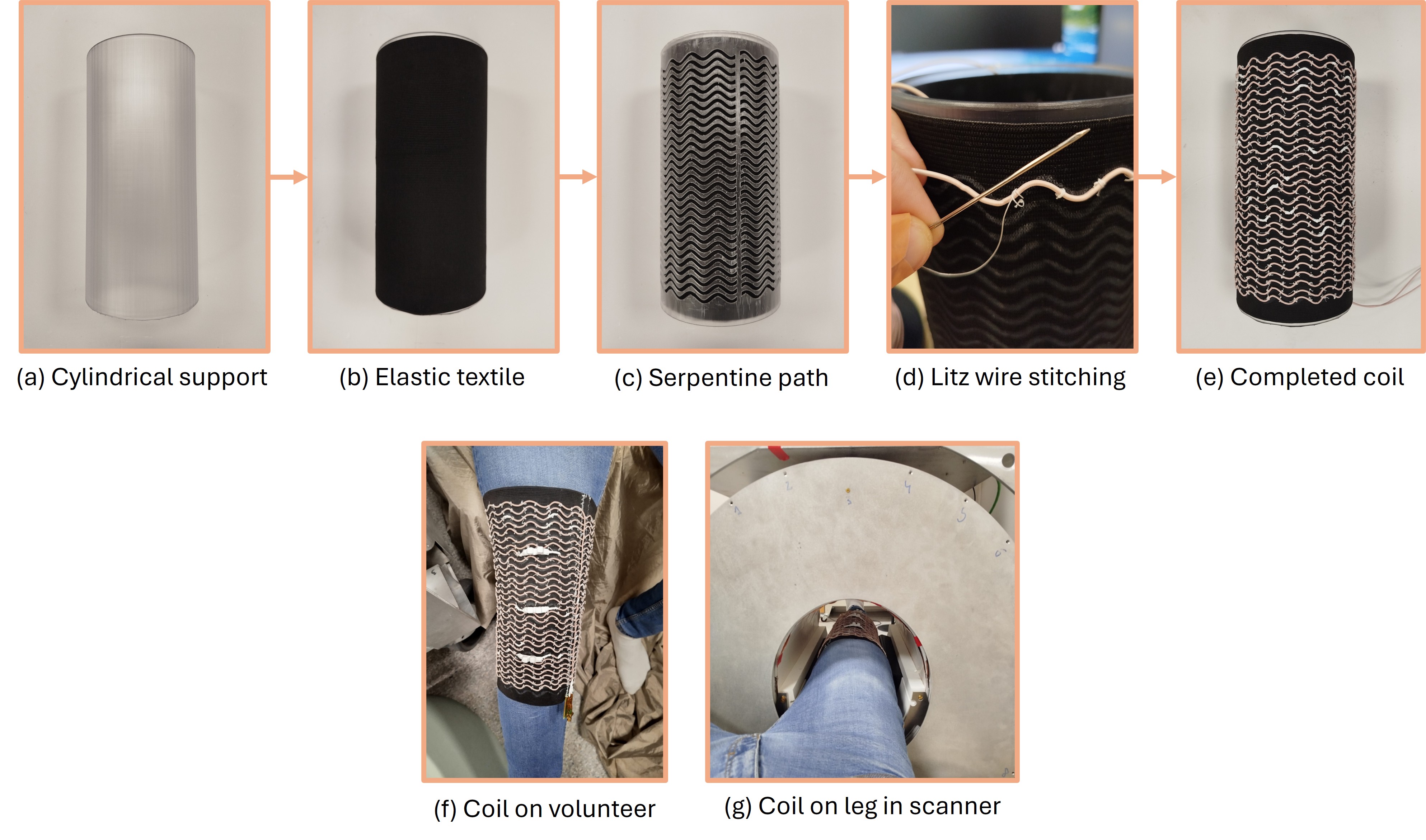}
    \caption{Fabrication of the stretchable RF coil. (a) 3D-printed cylindrical support. (b) Elastic textile placed around the cylindrical support. (c) 3D-printed guide defining the serpentine conductor path. (d) Litz wire hand-stitched onto the textile. (e) Completed coil. (f) Stretchable coil worn as a garment by the volunteer outside the MRI scanner, illustrating its close conformity to the lower-limb anatomy. (g) Stretchable coil positioned on the volunteer's leg inside the NextMRI scanner.}
    \label{fig:fabrication}
\end{figure}

The coil support was fabricated from an elastic textile manufactured by BENECREAT (China), comprising rubber covered with fibers, which allows it to be repeatedly stretched and subsequently recover its original shape. First, a cylindrical support was 3D printed in photopolymer resin, with an outer radius corresponding to the unstretched coil radius (Figure \ref{fig:fabrication}a). The elastic textile was sewn around this support (Figure \ref{fig:fabrication}b). A second 3D-printed component defined the serpentine path of the conductor (Figure \ref{fig:fabrication}c). The Litz wire was then manually stitched onto the textile at selected anchoring points that allowed the conductor path to deform during stretching (Figure \ref{fig:fabrication}d). The completed coil is shown in Figure \ref{fig:fabrication}e. The mechanical flexibility of the Litz wire facilitates deformation of the coil, whereas the serpentine conductor path provides the required circumferential stretchability. Figures \ref{fig:fabrication}f and \ref{fig:fabrication}g show the stretchable coil worn as a garment conforming to the volunteer's limb outside and inside the MRI scanner, respectively.

\subsubsection{Coil specifications}

The proposed wearable coil was compared with four rigid reference coils. Two rigid knee coils routinely used with the scanners served as the primary references. Both were wound using single-strand copper wire. The first had a diameter of 18\,cm, a length of 23\,cm, and 30 turns, and is hereafter denoted as $RF_\text{A}$. The second had a diameter of 15\,cm, a length of 23\,cm, and 44 turns, and is denoted as $RF_\text{B}$. The same nominal winding geometries were used in both MRI systems, although the copper-wire diameters differed between systems.

The stretchable coil was fabricated using the selected Litz wire. It had a length of 23\,cm and a diameter that could be increased from 13\,cm in its unstretched configuration to approximately 14.5\,cm, allowing adaptation to a wide range of lower-limb sizes. The coil comprised 30 turns. This number was selected as a compromise between the geometries of the reference coils and the spacing required to stitch the conductor onto the textile in a serpentine path while preserving its ability to stretch and recover its original shape. The resulting coil is hereafter denoted as $RF_\text{Stretch}$.

Two additional rigid Litz-wire coils were fabricated to independently evaluate the benefits of the Litz-wire conductor and the stretchable coil geometry. The first was a rigid replica of $RF_\text{B}$, denoted as $RF_\text{B-Litz}$, which differed from $RF_\text{B}$ only in the conductor material: Litz wire instead of conventional solid copper wire. This pair enabled an isolated assessment of the effect of the conductor material while maintaining the coil geometry.

\begin{figure}
    \centering
    \includegraphics[width=1\linewidth]{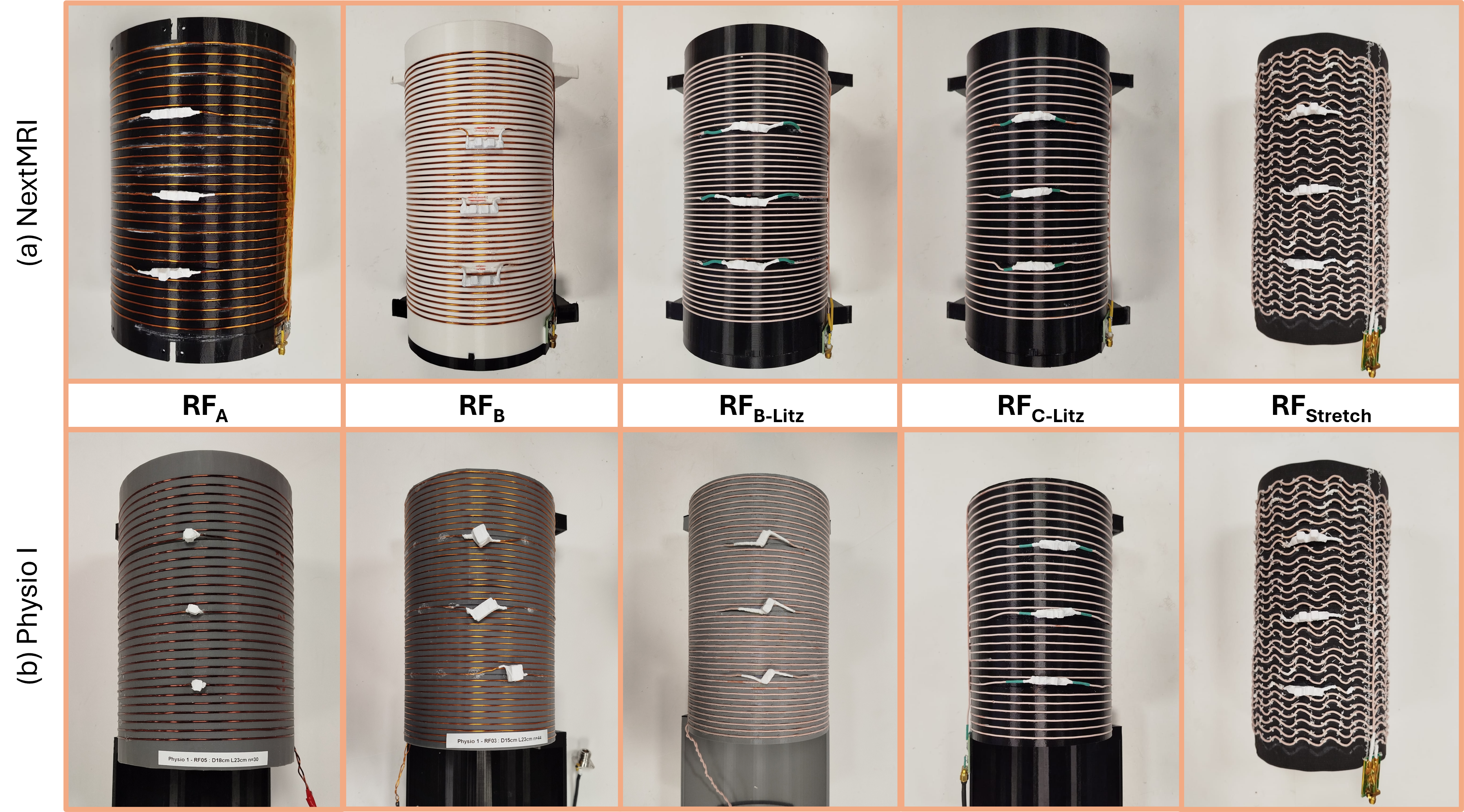}
    \caption{RF coils used for knee imaging with the (a) NextMRI and (b) Physio I scanners.}
    \label{fig:knee_coils}
\end{figure}

The second additional coil, denoted as $RF_\text{C-Litz}$, was a conventional rigid solenoid fabricated with Litz wire and a geometry closely matching that of $RF_\text{Stretch}$. It comprised 30 turns, had a length of 23\,cm, and had a diameter of 15\,cm. Because $RF_\text{C-Litz}$ and $RF_\text{Stretch}$ shared the same conductor material and closely matched geometries, their comparison enabled an isolated evaluation of the effects introduced by the stretchable implementation. Table \ref{tab:knee_coils_combined} summarizes the coil specifications, and Figures \ref{fig:knee_coils}a and \ref{fig:knee_coils}b show the five coils used with the NextMRI and the Physio I scanners, respectively.

\begin{table}
\centering
\renewcommand{\arraystretch}{1.2}
\begin{tabular}{lccccc}
\toprule
 & $RF_\text{A}$ & $RF_\text{B}$ & $RF_\text{B-Litz}$ &
 $RF_\text{C-Litz}$ & $RF_\text{Stretch}$ \\
\midrule
Coil length (mm)
    & 230 & 230 & 230 & 230 & 230 \\
Coil diameter (mm)
    & 180 & 150 & 150 & 150 & 130--145 \\
Number of turns
    & 30 & 44 & 44 & 30 & 30 \\
Wire diameter, NextMRI (mm)
    & 1.3 & 1.3 & 1.5 & 1.5 & 1.5 \\
Wire diameter, Physio I (mm)
    & 1.9 & 1.0 & 1.5 & 1.5 & 1.5 \\
Number of strands
    & 1 & 1 & 1{,}500 & 1{,}500 & 1{,}500 \\
Strand diameter ($\upmu$m)
    & -- & -- & 30 & 30 & 30 \\
\bottomrule
\end{tabular}
\caption{Specifications of the RF coils used for knee imaging with the NextMRI and Physio I scanners. The diameter of $RF_\text{Stretch}$ varies from 130\,mm in its unstretched configuration to approximately 145\,mm when stretched.}
\label{tab:knee_coils_combined}
\end{table}

\subsection{Experimental measurements}
\subsubsection{Litz-wire performance at different frequencies}
\label{subsec:litz_frequency}

To characterize the frequency-dependent electrical performance of the selected Litz wire, we compared two solenoids with identical geometries but different conductor materials: the solid-copper coil $RF_\text{B}$ and its Litz-wire counterpart, $RF_\text{B-Litz}$. This comparison isolated the effect of the conductor material and allowed us to evaluate how the electrical performance of the selected Litz wire varied with frequency \cite{webb2023tackling,giovannetti2017litz,pauly2006optimized}.

We measured the unloaded quality factor ($Q_\text{unloaded}$) and $B_1$ efficiency of both coils at 2.13\,MHz, 3.04\,MHz, 3.19\,MHz, 3.53\,MHz, 4.26\,MHz, and 5.32\,MHz, corresponding to magnetic field strengths of 50\,mT, 71.4\,mT, 75\,mT, 82.6\,mT, 100\,mT, and 125\,mT, respectively. A field strength of 50\,mT is representative of several low-field MRI systems \cite{o2021vivo,o2019three,obungoloch2023site,de2021design}, while other portable scanners operate at fields approaching or exceeding 100\,mT \cite{guallart2022portable,cooley2021portable,marques2019low,kimberly2023brain,hyperfine2023}. The frequencies corresponding to 71.4\,mT and 82.6\,mT were included because they match the operating conditions of Physio I and NextMRI, respectively. At each frequency, the gap capacitors were adjusted to tune the coil to resonance. All measurements were performed outside the scanners under identical experimental conditions, thereby avoiding differences associated with the shielding geometries of the two systems.

The unloaded quality factor was calculated as
\begin{equation}
    \label{eq:q1}
    Q_\text{unloaded} = \frac{f_0}{\Delta f_{-7\,\mathrm{dB}}},
\end{equation}
where $f_0$ is the resonant frequency and $\Delta f_{-7\,\mathrm{dB}}$ is the bandwidth between the two $-7\,\mathrm{dB}$ crossings of the $S_{11}$ magnitude response. For a resonator matched to 50\,\(\Omega\), these crossings correspond to the half-power points used to determine the quality factor \cite{Q2_kajfez1984q}. The resonant frequency and bandwidth were measured using a Rohde \& Schwarz FSH4 spectrum analyzer (Munich, Germany) configured as a vector network analyzer (VNA).

\begin{figure}
    \centering
    \includegraphics[width=0.7\linewidth]{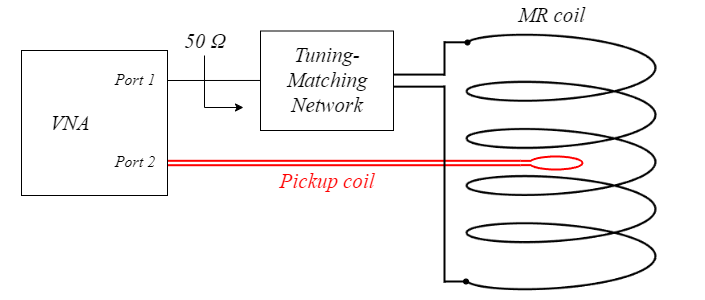}
    \caption{Simplified configuration used to estimate RF-coil efficiency from a transmission measurement with a pickup coil.}
    \label{fig:efficiencyS21_abbrev}
\end{figure}

The RF-coil efficiency was subsequently estimated from a transmission measurement performed using the same VNA. The coil under test was tuned to the corresponding frequency, matched to 50\,\(\Omega\), and connected to Port~1 of the VNA. A two-turn pickup coil with a diameter of 6.2\,mm and a wire diameter of 1\,mm was positioned at the center of the RF coil and connected to Port~2. Figure \ref{fig:efficiencyS21_abbrev} shows the measurement configuration.

The pickup-coil estimate of RF-coil efficiency was calculated as
\begin{equation}
    \label{eq:efficiency2}
    \varepsilon_\text{pickup} = \frac{B_1^+}{\sqrt{P_\text{in}}} = \frac{\sqrt{2R_\text{L}}}{4\pi fSN}  S_{21},
\end{equation}
where $B_1^+$ is the co-rotating component of the transmitted RF magnetic field, $P_\text{in}$ is the input power delivered to the RF coil, $R_\text{L}=50\,\Omega$ is the load resistance of the measurement system, $f$ is the operating frequency, $S$ is the area of the pickup coil, $N$ is its number of turns, and $S_{21}$ is the magnitude of the measured transmission coefficient expressed on a linear scale. The resulting efficiency is expressed in $\mathrm{T}/\sqrt{\mathrm{W}}$. %A derivation of Equation \ref{eq:efficiency2} is provided in Appendix \ref{an:efficiency_pickup}.

\subsubsection{Characterization of the RF coils inside the scanners}
\label{sec:RF_charac}
For each scanner, we measured the unloaded and loaded quality factors of all coils. Loaded measurements were performed using an American College of Radiology phantom (ACR; diameter, 10\,cm; depth, 10\,cm; filled with a NiCl$_2$/NaCl solution) and the right knee of the same volunteer for all coils. The loading factor (LF) was calculated as
\begin{equation}
    \text{LF} = 1-\frac{Q_\text{loaded}}{Q_\text{unloaded}},
\end{equation}
where $\text{LF}>0.5$ indicates that sample losses dominate over coil losses, whereas $\text{LF}<0.5$ indicates that coil losses remain dominant.

Before image acquisition, we also characterized the transmit efficiency of each coil inside the corresponding scanner using the ACR phantom and the volunteer's knee. The efficiency was determined from Rabi-flop measurements \cite{algarin2024marge}. Assuming that the coil was matched to 50\,\(\Omega\), the Rabi-based efficiency, $\varepsilon_\text{Rabi}$, was calculated as
\begin{equation}
    \label{eq:efficiency_rabi}
    \varepsilon_\text{Rabi} = \frac{B_1^+}{\sqrt{P_\text{in}}} = \frac{\pi}{\gamma t_{\pi}\sqrt{P_\text{in}}},
\end{equation}
where $B_1^+$ is the co-rotating component of the transmitted RF magnetic field, $P_\text{in}$ is the input power delivered to the coil, $\gamma$ is the gyromagnetic ratio of $^1$H, and $t_{\pi}$ is the duration of the $\pi$ pulse. The resulting efficiency is expressed in $\mathrm{T}/\sqrt{\mathrm{W}}$. 
%A derivation of Equation \ref{eq:efficiency_rabi} is provided in Appendix \ref{an:efficiency_rabi}.

\subsubsection{Imaging experiments}

Table \ref{tab:sequence_parameters} summarizes the acquisition parameters of the RARE sequences used for phantom and \emph{in-vivo} imaging \cite{hennig1986rare}. Experiments I and II comprised phantom imaging with NextMRI and Physio I, respectively. Experiment I consisted of transverse T$_1$-weighted imaging of the ACR phantom with NextMRI, whereas Experiment II consisted of sagittal proton-density-weighted imaging with Physio I.

Experiments III and IV comprised \emph{in-vivo} imaging with NextMRI and Physio I, respectively. In both experiments, sagittal T$_1$-weighted images of the right knee of the same volunteer were acquired using all five RF coils. Additional imaging protocols were performed using $RF_\text{Stretch}$ and the rigid coils with the highest and lowest SNR in each scanner. In NextMRI, these were $RF_\text{C-Litz}$ and $RF_\text{A}$, respectively; in Physio I, they were $RF_\text{B-Litz}$ and $RF_\text{A}$, respectively. The NextMRI protocol included sagittal T$_1$-weighted and T$_2$-weighted acquisitions and short-tau inversion recovery (STIR) acquisitions in the sagittal and axial planes. The Physio I protocol included sagittal T$_1$-weighted, T$_2$-weighted, and STIR acquisitions. Within each scanner and image contrast, identical acquisition parameters were used for all RF coils.

\begin{table}
\centering
\scriptsize
\setlength{\tabcolsep}{2pt}
\renewcommand{\arraystretch}{1.15}
\begin{tabular}{@{}c c c c c c c c c c c c@{}}
\toprule
\makecell{Exp.\\(figure)} &
Contrast &
Plane &
\makecell{FOV\\(mm$^3$)} &
Matrix &
\makecell{BW\\(kHz)} &
\makecell{TI/TE/TR\\(ms)} &
ETL &
\makecell{PF\\(\%)} &
$N$ &
\makecell{Time\\(min)} &
RF coils \\
\midrule

\makecell{I\\Fig.~\ref{fig:img_phantom}a} &
T$_1$-w &
tra &
$150\times150\times150$ &
$320\times320\times36$ &
26.7 &
--/20/200 &
5 &
100 &
1 &
7.7 &
All
\\

\makecell{II\\Fig.~\ref{fig:img_phantom}b} &
PD-w &
sag &
$140\times140\times140$ &
$120\times120\times120$ &
30.0 &
--/10/800 &
5 &
65 &
1 &
25.0 &
All
\\

\makecell{III\\Figs.~\ref{fig:img_knee}a,\\\ref{fig:processed_next}a} &
T$_1$-w &
sag &
$180\times160\times160$ &
$226\times200\times30$ &
37.6 &
--/20/200 &
5 &
70 &
2 &
5.6 &
All
\\

\makecell{III\\Fig.~\ref{fig:processed_next}b} &
T$_2$-w &
sag &
$180\times160\times160$ &
$168\times150\times30$ &
28.0 &
--/20/1000 &
20 &
60 &
2 &
4.5 &
\makecell{$RF_\text{A}$, $RF_\text{C-Litz}$,\\$RF_\text{Stretch}$}
\\

\makecell{III\\Figs.~\ref{fig:processed_next}c,\\\ref{fig:processed_next}d} &
STIR &
sag, ax &
$180\times160\times160$ &
$136\times120\times30$ &
34.0 &
90/20/1000 &
5 &
60 &
2 &
14.4 &
\makecell{$RF_\text{A}$, $RF_\text{C-Litz}$,\\$RF_\text{Stretch}$}
\\

\makecell{IV\\Figs.~\ref{fig:img_knee}b,\\\ref{fig:processed_physio}a} &
T$_1$-w &
sag &
$240\times180\times160$ &
$160\times120\times32$ &
40.0 &
--/10/150 &
6 &
65 &
10 &
10.4 &
All
\\

\makecell{IV\\Fig.~\ref{fig:processed_physio}b} &
T$_2$-w &
sag &
$240\times180\times160$ &
$160\times120\times32$ &
53.3 &
--/10/1000 &
20 &
65 &
5 &
10.4 &
\makecell{$RF_\text{A}$, $RF_\text{B-Litz}$,\\$RF_\text{Stretch}$}
\\

\makecell{IV\\Fig.~\ref{fig:processed_physio}c} &
STIR &
sag &
$240\times180\times160$ &
$160\times120\times32$ &
40.0 &
65/10/800 &
20 &
65 &
6 &
10.0 &
\makecell{$RF_\text{A}$, $RF_\text{B-Litz}$,\\$RF_\text{Stretch}$}
\\

\bottomrule
\end{tabular}
\caption{Acquisition parameters of the RARE sequences used for phantom and \emph{in-vivo} imaging. Timing parameters are reported as TI/TE/TR; ``--'' indicates that no inversion pulse was applied. BW, acquisition bandwidth; ETL, echo-train length; FOV, field of view; $N$, number of scans; PD-w, proton-density-weighted; PF, partial Fourier; STIR, short-tau inversion recovery; ax, axial; sag, sagittal; tra, transverse.}
\label{tab:sequence_parameters}
\end{table}

\subsubsection{SNR analysis}
\label{subsec:SNRanalysis}

The SNR was estimated from the raw reconstructed images using dedicated noise acquisitions. For each imaging experiment, noise-only data were acquired immediately before image acquisition with RF excitation disabled, while maintaining the same hardware and receiver settings. This approach provided a direct estimate of the system noise without contamination from residual MR signal or image artifacts \cite{kellman2005image}.

The reconstructed magnitude image volume, $I(x,y,z)$, was normalized by the image-domain noise standard deviation, $\sigma_\text{img}$, yielding a dimensionless voxel-wise SNR. The image-domain noise standard deviation was estimated from the standard deviation of the complex samples obtained during the corresponding noise-only acquisition, $\sigma_\text{noise}$, accounting for the normalization introduced by the three-dimensional inverse Fourier transform and the effective number of acquired k-space samples:
\begin{equation}
    \sigma_\text{img} = \frac{\sigma_\text{noise}}{\sqrt{N_xN_yN_z}\sqrt{\text{PF}}},
\end{equation}
where $N_x$, $N_y$, and $N_z$ are the reconstructed matrix dimensions along the readout, phase-encoding, and slice-encoding directions, respectively, and PF is the partial Fourier factor applied along the second phase-encoding direction, expressed as a fraction between 0 and 1. The voxel-wise SNR was therefore calculated as
\begin{equation}
    \mathrm{SNR}(x,y,z) = \frac{I(x,y,z)}{\sigma_\text{img}} = \frac{I(x,y,z)\sqrt{N_xN_yN_z}\sqrt{\text{PF}}}{\sigma_\text{noise}}.
\end{equation}

This procedure follows the principle that reconstructed images can be expressed directly in SNR units when the acquisition noise statistics and reconstruction scaling are known \cite{kellman2005image}. It avoids the bias that can arise when estimating noise from background regions of magnitude MR images, where the nonlinear magnitude reconstruction alters the underlying noise distribution and introduces a noise-dependent bias, particularly at low signal levels \cite{henkelman1985measurement,sijbers1998estimation}.

For visualization, the SNR maps were smoothed using Block-Matching and Four-Dimensional Collaborative Filtering (BM4D) \cite{bm4d_maggioni2012nonlocal}. This filtering was applied only to the displayed SNR maps and did not affect the quantitative analysis, which was performed using the raw unfiltered maps.

For each image, the global SNR was calculated as the mean SNR across all pixels. SNR was also evaluated within two representative areas, denoted A and B, which are identified in the corresponding figures. For comparison among coil designs, each SNR measurement was normalized to that obtained with the reference coil, $RF_\text{A}$, which is the standard coil used for patient imaging in both systems:
\begin{equation}
    \mathrm{SNR}_{\text{rel},i} = \frac{\mathrm{SNR}_i}{\mathrm{SNR}_{RF_\text{A}}},
\end{equation}
such that $\mathrm{SNR}_{\text{rel},i}>1$ indicates an improvement relative to $RF_\text{A}$.

\subsubsection{Image post-processing}
\label{sec:img_post}
The complete knee imaging protocols were post-processed using a deep-learning denoising algorithm based on a pre-trained high-field model developed with the SNRAware methodology \cite{xue2025snraware,guallart2026high}. Geometric distortions caused by $B_0$ inhomogeneity were corrected using the Single-Point Double-Shot (SPDS) method \cite{SPDS_borreguero2025zero}. Following distortion correction, approximately 30 to 35\,\% of the pixels were removed from the peripheral regions of each image to retain the diagnostically relevant field of view and exclude residual distortion artifacts.

This post-processing pipeline was applied to the datasets acquired in Experiments III and IV using the three RF coils selected for each scanner: $RF_\text{Stretch}$, the rigid coil with the highest SNR, and the rigid coil with the lowest SNR. The corresponding acquisition parameters are provided in Table \ref{tab:sequence_parameters}.

%%%%%%%%%%%%%%%%%%%%%%%% RESULTS %%%%%%%%%%%%%%%%%%%%%%%%%%%
    
\section{Results}
\subsection{Evaluation of Litz-wire performance at different frequencies}
\begin{figure}
    \centering
    \includegraphics[width=0.65\linewidth]{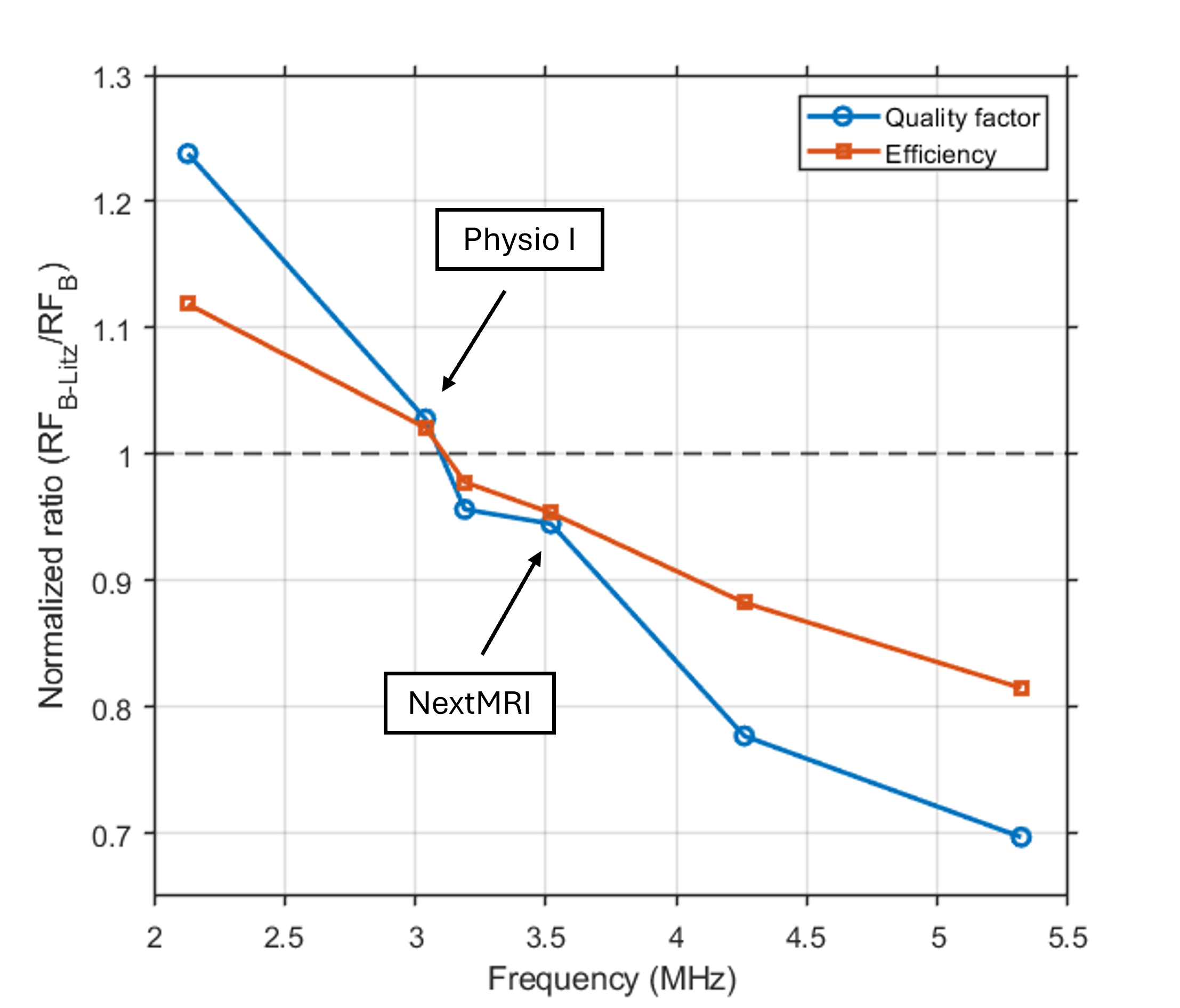}
    \caption{Ratios of the unloaded quality factor and pickup-coil efficiency measured for $RF_\text{B-Litz}$ and $RF_\text{B}$.}
    \label{fig:Q_eff_freqs}
\end{figure}

Table \ref{tab:Q_eff_freqs} displays the $Q_\text{unloaded}$ and $\varepsilon_\text{pickup}$ values measured for $RF_\text{B}$ and $RF_\text{B-Litz}$ over the investigated frequency range using the procedure described in Section \ref{subsec:litz_frequency}. Figure \ref{fig:Q_eff_freqs} shows the corresponding ratios between the Litz-wire and solid-copper coils for both parameters.

\begin{table}
\centering
\caption{Unloaded quality factor, $Q_\text{unloaded}$, and pickup-coil efficiency, $\varepsilon_\text{pickup}$, measured for $RF_\text{B}$ and $RF_\text{B-Litz}$ as a function of frequency.}
\renewcommand{\arraystretch}{1.2}
\begin{tabularx}{0.65\textwidth}{cYYYY}
\toprule
\multirow{2}{*}{$f_0$ (MHz)} &
\multicolumn{2}{c}{$Q_\text{unloaded}$} &
\multicolumn{2}{c}{$\varepsilon_\text{pickup}$ ($\upmu\mathrm{T}/\sqrt{\mathrm{W}}$)} \\
\cmidrule(lr){2-3}
\cmidrule(lr){4-5}
& $RF_\text{B}$ & $RF_\text{B-Litz}$
& $RF_\text{B}$ & $RF_\text{B-Litz}$ \\
\midrule
2.13 & 517 & 640 & 59 & 66 \\
3.04 & 504 & 518 & 48 & 49 \\
3.19 & 457 & 437 & 45 & 44 \\
3.53 & 435 & 411 & 43 & 41 \\
4.26 & 363 & 282 & 34 & 30 \\
5.32 & 277 & 193 & 27 & 22 \\
\bottomrule
\end{tabularx}
\label{tab:Q_eff_freqs}
\end{table}

%%%%%%%%%%%%%%%%%%%%%%%% Q-factor and Efficiency Results %%%%%%%%%%%%%%%%%%%%%%%%%%%%%
\subsection{Characterization of the RF coils inside the scanners}

Table \ref{tab:Q_RFs} displays the unloaded and loaded quality factors measured for all RF coils, together with the corresponding loading factors under ACR phantom and \emph{in-vivo} knee loading. Table \ref{tab:eff_RFs} displays the transmit efficiencies measured for all RF coils using the Rabi-flop method before imaging the ACR phantom and the knee of the same volunteer in both scanners. Measurements were performed as described in Sec.~\ref{sec:RF_charac}.

\begin{table}
\centering
\caption{Unloaded and loaded quality factors and corresponding loading factors for all RF coils under ACR phantom and \emph{in-vivo} knee loading. The highest $Q$-values for each load appear in bold.}
\label{tab:Q_RFs}
\renewcommand{\arraystretch}{1.2}
\small
\begin{tabularx}{0.9\textwidth}{llcYYYYY}
\toprule
Scanner &
Load &
Metric &
$RF_\text{A}$ &
$RF_\text{B}$ &
$RF_\text{B-Litz}$ &
$RF_\text{C-Litz}$ &
$RF_\text{Stretch}$ \\
\midrule

\multirow{5}{*}{NextMRI}
& None        & $Q_\text{unloaded}$ & 412 & 444 & 404 & 530 & \textbf{463} \\
& ACR phantom & $Q_\text{loaded}$   & 353 & 327 & 309 & \textbf{383} & 336 \\
&             & $\text{LF}$         & 0.14 & 0.26 & 0.24 & 0.28 & 0.27 \\
& Knee        & $Q_\text{loaded}$   & \textbf{285} & 218 & 202 & 247 & 128 \\
&             & $\text{LF}$         & 0.31 & 0.51 & 0.50 & 0.53 & 0.72 \\
\midrule

\multirow{5}{*}{Physio I}
& None        & $Q_\text{unloaded}$ & 312 & 223 & \textbf{563} & 493 & 457 \\
& ACR phantom & $Q_\text{loaded}$   & 282 & 192 & \textbf{436} & 383 & 369 \\
&             & $\text{LF}$         & 0.10 & 0.14 & 0.23 & 0.22 & 0.19 \\
& Knee        & $Q_\text{loaded}$   & 252 & 99  & 208 & \textbf{274} & 110 \\
&             & $\text{LF}$         & 0.19 & 0.56 & 0.63 & 0.44 & 0.76 \\

\bottomrule
\end{tabularx}
\end{table}

\begin{table}
\centering
\caption{Rabi-based transmit efficiency, $\varepsilon_\text{Rabi}$, measured for all RF coils under ACR phantom and \emph{in-vivo} knee loading. Values are expressed in $\upmu\mathrm{T}/\sqrt{\mathrm{W}}$. The highest values for each load appear in bold.}
\label{tab:eff_RFs}
\renewcommand{\arraystretch}{1.2}
\begin{tabularx}{0.85\textwidth}{llYYYYY}
\toprule
Scanner &
Load &
$RF_\text{A}$ &
$RF_\text{B}$ &
$RF_\text{B-Litz}$ &
$RF_\text{C-Litz}$ &
$RF_\text{Stretch}$ \\
\midrule

\multirow{2}{*}{NextMRI}
& ACR phantom & 26 & 34 & 33 & 37 & \textbf{40} \\
& Knee        & 24 & 26 & 28 & 30 & \textbf{31} \\
\midrule

\multirow{2}{*}{Physio I}
& ACR phantom & 27 & 36 & 55 & 50 & \textbf{56} \\
& Knee        & 27 & 32 & 46 & 43 & \textbf{48} \\

\bottomrule
\end{tabularx}
\end{table}

\subsection{Phantom imaging}

\begin{figure}
    \centering
    \includegraphics[width=0.8\linewidth]{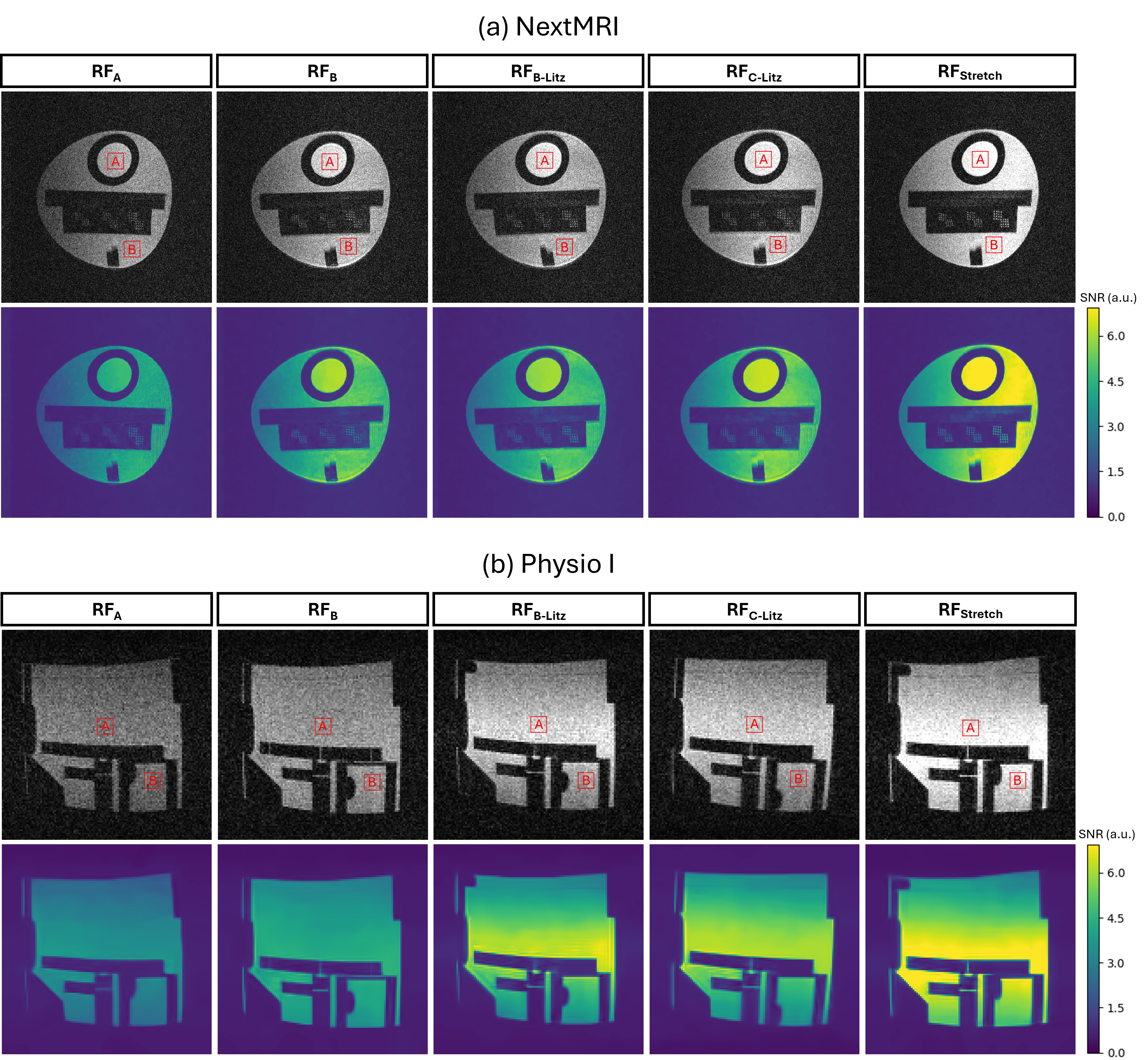}
    \caption{ACR phantom images and corresponding SNR maps acquired with the five RF coils in (a) NextMRI and (b) Physio I.}
    \label{fig:img_phantom}
\end{figure}

Figures \ref{fig:img_phantom}a and \ref{fig:img_phantom}b show the ACR phantom images acquired with the five RF coils in NextMRI and Physio I, respectively, together with the corresponding SNR maps. A common intensity scale was used for all anatomical images, and a common SNR scale was used for all SNR maps. Table \ref{tab:img_phantom} reports the absolute and relative mean SNR calculated within regions A and B and across the entire image. It also reports the RMS noise level relative to the theoretical Johnson noise \cite{webb2023tackling,guallart2026electromagnetic}. Regions A and B were defined using square ROIs of $24\times24$ pixels in the NextMRI images and $9\times9$ pixels in the Physio I images, corresponding to approximately the same physical area.

\begin{table}
\centering
\caption{Absolute and relative mean SNR calculated from the raw ACR phantom images acquired with NextMRI and Physio I. RMS noise is reported relative to the theoretical Johnson noise. The highest SNR-values appear in bold.}
\label{tab:img_phantom}
\renewcommand{\arraystretch}{1.2}
\small
\begin{tabularx}{\textwidth}{llcYYYYY}
\toprule
Scanner &
Measurement &
Metric &
$RF_\text{A}$ &
$RF_\text{B}$ &
$RF_\text{B-Litz}$ &
$RF_\text{C-Litz}$ &
$RF_\text{Stretch}$ \\
\midrule

\multirow{7}{*}{NextMRI}
& SNR A      & Absolute & 4.68 & 6.12 & 5.94 & 6.54 & \textbf{7.55} \\
&            & Relative & 1.00 & 1.31 & 1.27 & 1.40 & \textbf{1.61} \\
\cmidrule(lr){2-8}
& SNR B      & Absolute & 4.48 & 5.31 & 5.09 & 5.72 & \textbf{6.96} \\
&            & Relative & 1.00 & 1.19 & 1.13 & 1.28 & \textbf{1.55} \\
\cmidrule(lr){2-8}
& Global SNR & Absolute & 1.25 & 1.45 & 1.42 & 1.49 & \textbf{1.60} \\
&            & Relative & 1.00 & 1.16 & 1.13 & 1.19 & \textbf{1.28} \\
\cmidrule(lr){2-8}
& RMS noise  & Johnson-noise ratio & 1.20 & 1.19 & 1.19 & 1.19 & 1.19 \\
\midrule

\multirow{7}{*}{Physio I}
& SNR A      & Absolute & 3.39 & 4.14 & 5.97 & 5.77 & \textbf{6.82} \\
&            & Relative & 1.00 & 1.22 & 1.76 & 1.70 & \textbf{2.01} \\
\cmidrule(lr){2-8}
& SNR B      & Absolute & 2.98 & 4.12 & 4.47 & 4.27 & \textbf{6.36} \\
&            & Relative & 1.00 & 1.38 & 1.50 & 1.43 & \textbf{2.13} \\
\cmidrule(lr){2-8}
& Global SNR & Absolute & 1.16 & 1.41 & 1.59 & 1.51 & \textbf{1.90} \\
&            & Relative & 1.00 & 1.22 & 1.37 & 1.31 & \textbf{1.64} \\
\cmidrule(lr){2-8}
& RMS noise  & Johnson-noise ratio & 1.24 & 1.29 & 1.25 & 1.25 & 1.32 \\

\bottomrule
\end{tabularx}
\end{table}

\subsection{\emph{In-vivo} imaging}
\begin{figure}
    \centering
    \includegraphics[width=0.9\linewidth]{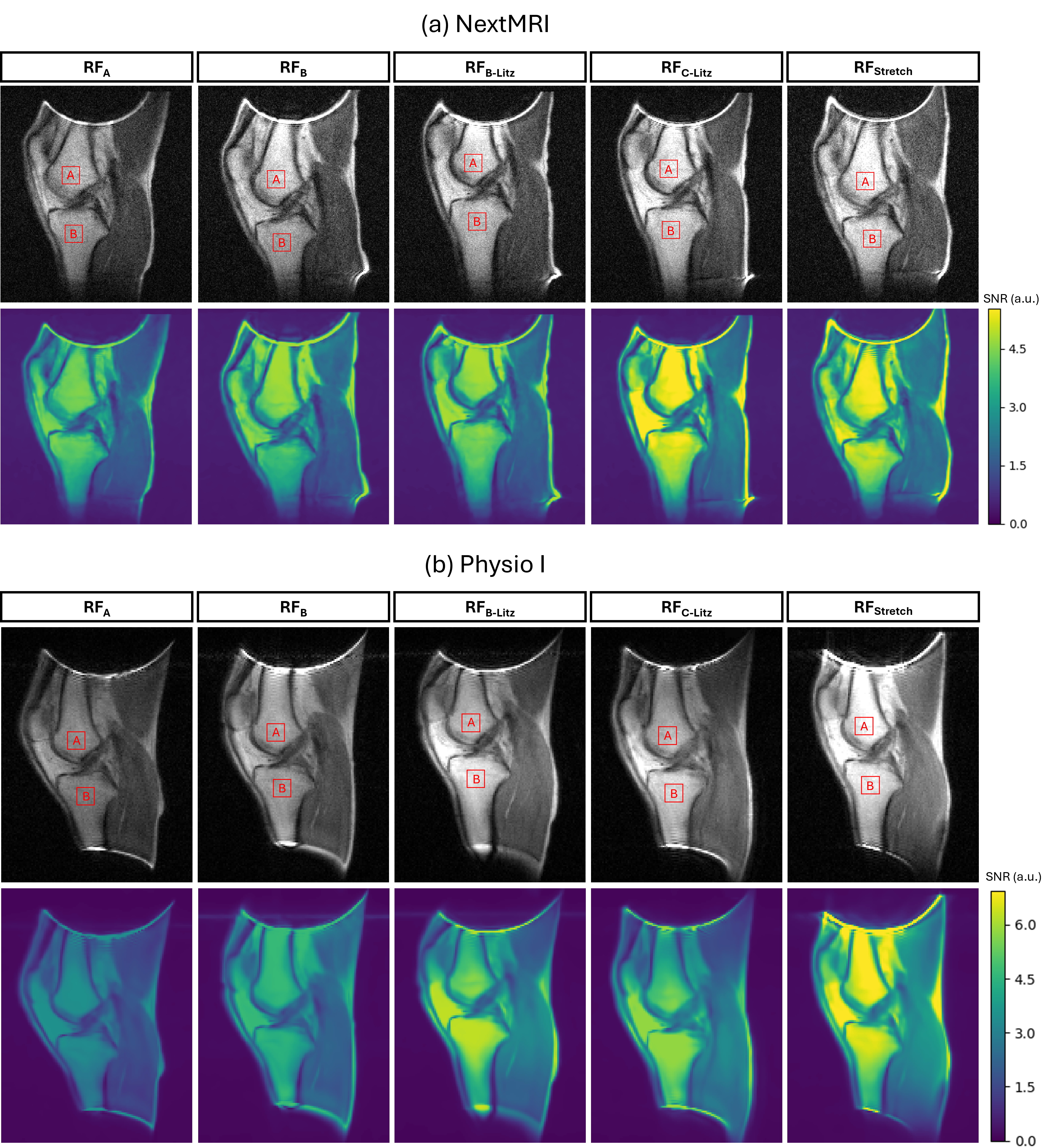}
    \caption{T$_1$-weighted knee images and corresponding SNR maps acquired from the same volunteer using the five RF coils in (a) NextMRI and (b) Physio I.}
    \label{fig:img_knee}
\end{figure}

Figures \ref{fig:img_knee}a and \ref{fig:img_knee}b show the T$_1$-weighted knee images acquired from the same volunteer using the five RF coils in NextMRI and Physio I, respectively, together with the corresponding SNR maps. A common intensity scale was used for all anatomical images, and a common SNR scale was used for all SNR maps. Table \ref{tab:img_knee} reports the absolute and relative mean SNR calculated within regions A and B and across the entire image. Regions A and B were defined manually using square ROIs positioned over the distal femoral condyle and proximal tibial plateau, respectively. ROI dimensions were fixed within each scanner at $18\times18$ pixels for NextMRI and $11\times11$ pixels for Physio I, corresponding to approximately the same physical area.

\begin{table}
\centering
\caption{Absolute and relative mean SNR calculated from the raw \emph{in-vivo} knee images acquired with NextMRI and Physio I. RMS noise is reported relative to the theoretical Johnson noise. The highest SNR-values appear in bold.}
\label{tab:img_knee}
\renewcommand{\arraystretch}{1.2}
\small
\begin{tabularx}{\textwidth}{llcYYYYY}
\toprule
Scanner &
Measurement &
Metric &
$RF_\text{A}$ &
$RF_\text{B}$ &
$RF_\text{B-Litz}$ &
$RF_\text{C-Litz}$ &
$RF_\text{Stretch}$ \\
\midrule

\multirow{7}{*}{NextMRI}
& SNR A      & Absolute & 4.40 & 4.53 & 4.62 & \textbf{5.54} & 5.33 \\
&            & Relative & 1.00 & 1.03 & 1.05 & \textbf{1.26} & 1.21 \\
\cmidrule(lr){2-8}
& SNR B      & Absolute & 3.99 & 3.68 & 4.32 & \textbf{5.12} & 4.89 \\
&            & Relative & 1.00 & 0.92 & 1.09 & \textbf{1.28} & 1.23 \\
\cmidrule(lr){2-8}
& Global SNR & Absolute & 0.91 & 0.96 & 0.94 & \textbf{1.13} & 1.07 \\
&            & Relative & 1.00 & 1.06 & 1.04 & \textbf{1.24} & 1.18 \\
\cmidrule(lr){2-8}
& RMS noise  & Johnson-noise ratio & 1.22 & 1.36 & 1.36 & 1.27 & 1.39 \\
\midrule

\multirow{7}{*}{Physio I}
& SNR A      & Absolute & 3.70 & 4.45 & 5.72 & 5.56 & \textbf{7.01} \\
&            & Relative & 1.00 & 1.20 & 1.55 & 1.50 & \textbf{1.89} \\
\cmidrule(lr){2-8}
& SNR B      & Absolute & 3.37 & 4.28 & 6.36 & 6.02 & \textbf{6.72} \\
&            & Relative & 1.00 & 1.27 & 1.89 & 1.79 & \textbf{1.99} \\
\cmidrule(lr){2-8}
& Global SNR & Absolute & 0.68 & 0.83 & 1.02 & 1.07 & \textbf{1.22} \\
&            & Relative & 1.00 & 1.23 & 1.51 & 1.58 & \textbf{1.80} \\
\cmidrule(lr){2-8}
& RMS noise  & Johnson-noise ratio & 1.42 & 1.37 & 1.34 & 1.26 & 1.30 \\

\bottomrule
\end{tabularx}
\end{table}

\subsection{Image post-processing}
\begin{figure}
    \centering
    \includegraphics[width=\linewidth]{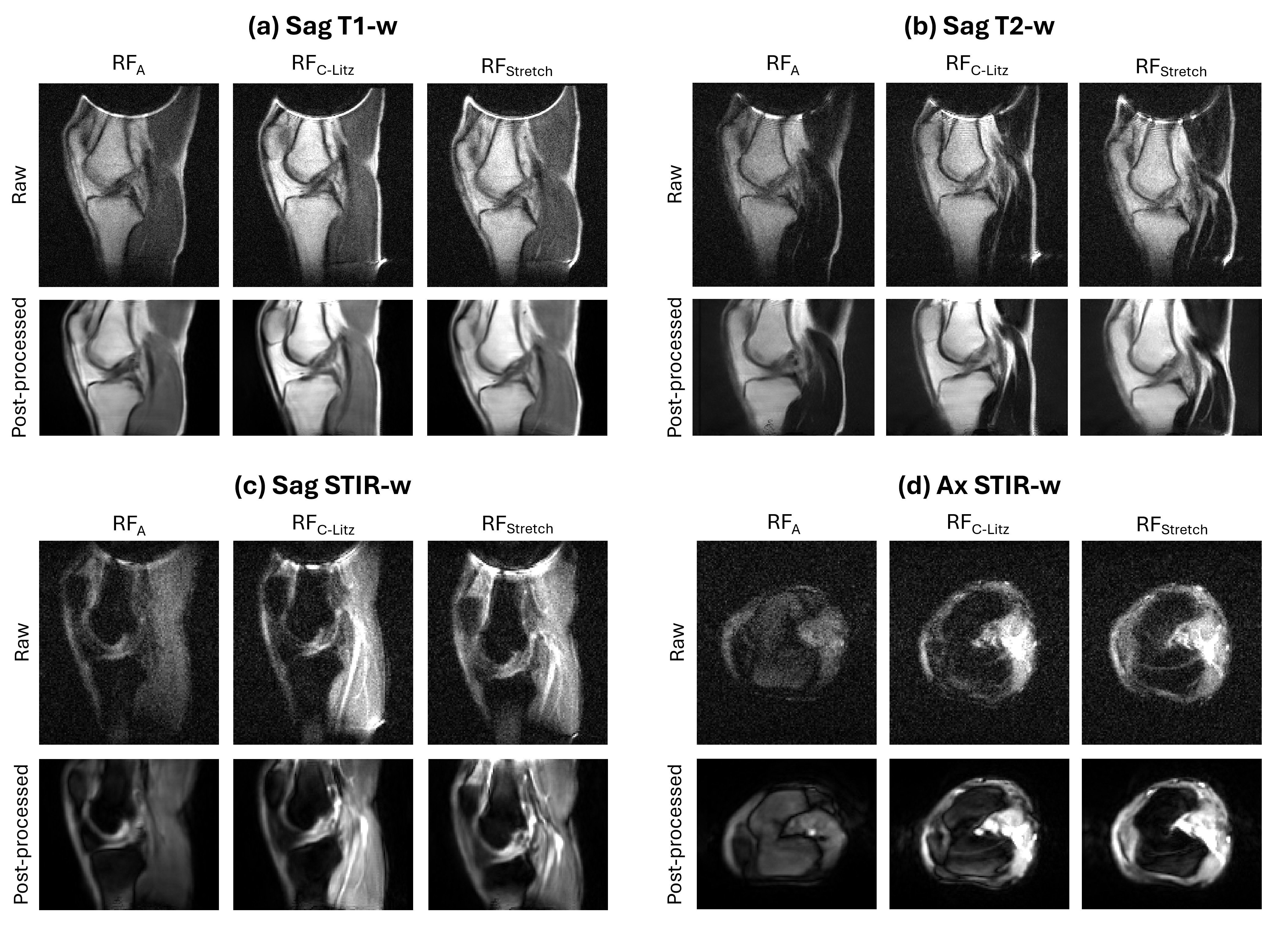}
    \caption{Knee images acquired in NextMRI before and after post-processing. (a) Sagittal T$_1$-weighted images. (b) Sagittal T$_2$-weighted images. (c) Sagittal STIR images. (d) Axial STIR images.}
    \label{fig:processed_next}
\end{figure}

Figure \ref{fig:processed_next} shows the knee images from Experiment III (Sec.~\ref{sec:img_post}) before and after application of the SNRAware denoising pipeline, SPDS distortion correction, and image cropping. Images were acquired in NextMRI using $RF_\text{Stretch}$, the rigid coil with the highest SNR ($RF_\text{C-Litz}$), and the rigid coil with the lowest SNR ($RF_\text{A}$).

\begin{figure}
    \centering
    \includegraphics[width=\linewidth]{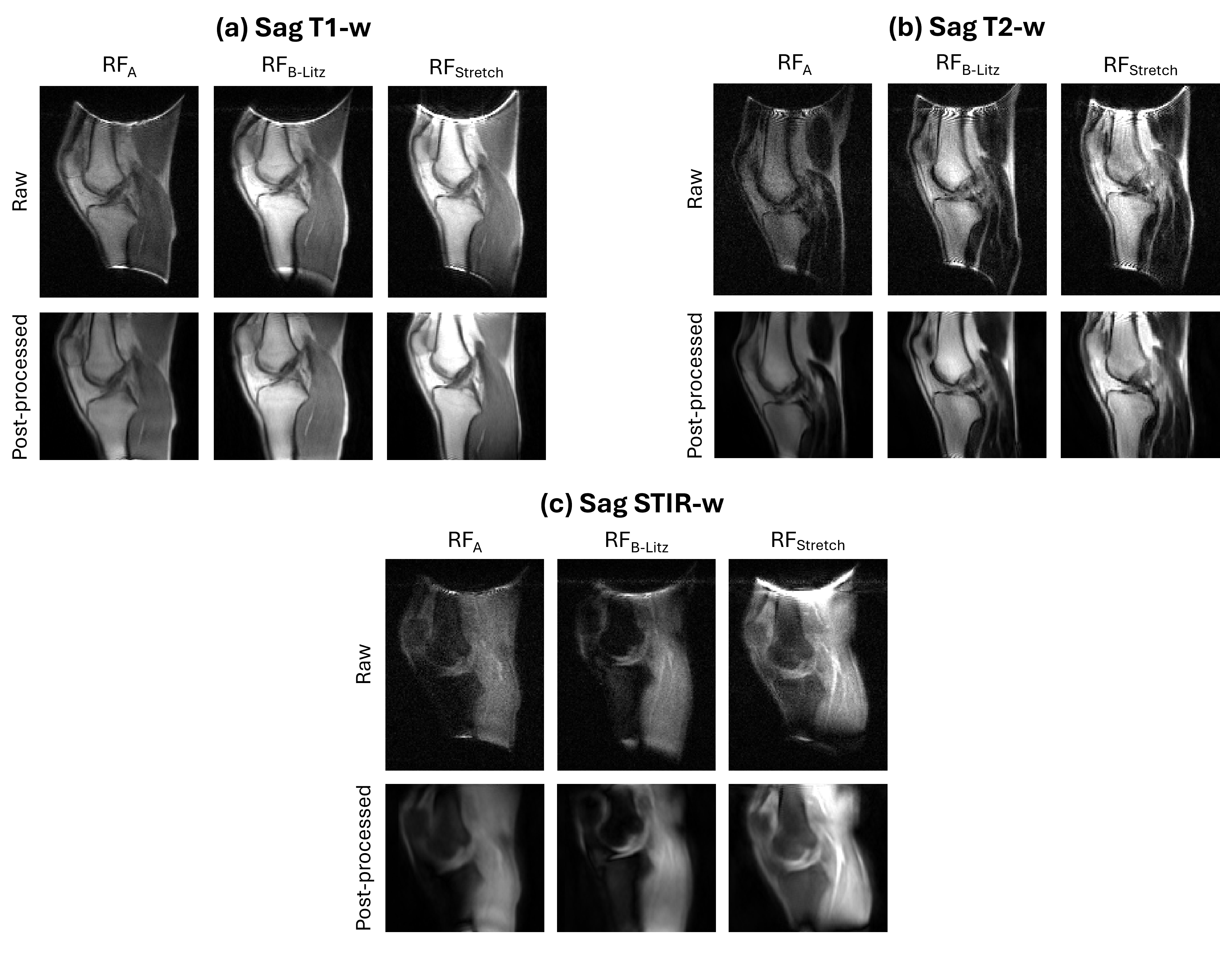}
    \caption{Knee images acquired in Physio I before and after post-processing. (a) Sagittal T$_1$-weighted images. (b) Sagittal T$_2$-weighted images. (c) Sagittal STIR images.}
    \label{fig:processed_physio}
\end{figure}

Figure \ref{fig:processed_physio} shows the corresponding knee images from Experiment IV before and after application of the same post-processing pipeline. Again, images were acquired in Physio I using $RF_\text{Stretch}$, the rigid coil with the highest SNR ($RF_\text{B-Litz}$), and the rigid coil with the lowest SNR ($RF_\text{A}$).

%%%%%%%%%%%%%%%%%%%%%%%%%%%%%%%% DISCUSSION %%%%%%%%%%%%%%%%%%%%%%%%%%%%%%%

\section{Discussion}

The performance of the proposed wearable RF coil is governed by two complementary mechanisms: the frequency-dependent electrical behavior of the Litz-wire conductor and the greater anatomical conformity enabled by the stretchable implementation. The respective contributions of these mechanisms, together with the practical advantages of the proposed design, are discussed below.

\subsection{Frequency-dependent performance of the selected Litz wire}

The results presented in Table \ref{tab:Q_eff_freqs} and Figure \ref{fig:Q_eff_freqs} characterize the frequency-dependent electrical performance of the selected Litz wire, which was chosen from among several Litz-wire configurations based on both their electrical performance at the operating frequencies and their mechanical suitability for the intended coil geometry. Its advantage over solid copper progressively decreases with increasing frequency, with a crossover at approximately 3.1\,MHz. At the Physio I operating frequency of 3.04\,MHz, the selected Litz wire provides slightly higher $Q_\text{unloaded}$ and $\varepsilon_\text{pickup}$ than the solid-copper conductor. In contrast, at the NextMRI operating frequency of 3.53\,MHz, it results in slightly lower values for both metrics. Although the exact values may vary slightly depending on the diameter of the solid-copper conductor used for comparison, these results are representative of the relative frequency-dependent performance of the two conductor types.

This frequency-dependent behavior is also observed in the measurements performed inside the MRI scanners. Across the quality factor, $\varepsilon_\text{rabi}$, phantom SNR, and \emph{in-vivo} SNR measurements, $RF_{B-Litz}$ consistently outperforms $RF_B$ in Physio I, with a more pronounced difference between the two coils. In NextMRI, the difference is substantially smaller, with $RF_{B-Litz}$ generally showing slightly lower performance than $RF_B$, although the relative performance depends on the measured metric. These results further support the frequency-dependent nature of the electrical benefit provided by the selected Litz wire under realistic MRI operating conditions.

Conductor selection for wearable RF coils cannot, however, be based exclusively on electrical performance. Although the selected Litz wire is slightly disadvantageous at 3.53\,MHz, its flexibility facilitates textile integration and repeated mechanical deformation. As discussed in the following sections, these mechanical properties enable a stretchable implementation whose greater anatomical conformity can compensate for, or outweigh, the modest electrical penalty through an increased filling factor.

\subsection{Effect of stretchability, anatomical conformity, and loading}

The effect of the stretchable implementation can be assessed by comparing $RF_\text{C-Litz}$ and $RF_\text{Stretch}$, which use the same conductor material and have similar electromagnetic designs. Although their geometries are not strictly identical, their primary difference is the ability of $RF_\text{Stretch}$ to conform to the anatomy. This conformity reduces the separation between the coil and the sample, thereby increasing the filling factor and strengthening their electromagnetic coupling.

The loading measurements in Table \ref{tab:Q_RFs} support this interpretation. Loaded with the ACR phantom, both coils exhibit similar loading factors in both MRI systems, since $RF_\text{Stretch}$ is basically undeformed. \emph{In vivo}, however, $RF_\text{Stretch}$ exhibits substantially higher loading factors than $RF_\text{C-Litz}$: 0.72 versus 0.53 in NextMRI and 0.76 versus 0.44 in Physio I. These values indicate that $RF_\text{Stretch}$ operates in a sample-noise-dominated regime in both systems. By comparison, $RF_\text{C-Litz}$ remains close to the transition between coil- and sample-noise dominance in NextMRI and is coil-noise dominated in Physio I. 

This behavior is consistent with the stronger coupling produced by the greater anatomical conformity and increased filling factor of the wearable coil. A higher loading factor corresponds to a greater reduction in $Q$ upon loading and therefore indicates stronger electrical coupling between the coil and the sample. Such electrical coupling is not necessarily desirable, as capacitive coupling between the body and the RF coil can also provide a pathway for electromagnetic interference in low-field MRI \cite{pfitzer2026fence}. Therefore, the loading factor should not be interpreted as an intrinsic measure of coil performance. Instead, it provides a means of comparing the degree of sample coupling achieved by the wearable coil relative to the rigid coils, while RF performance is evaluated separately through coil efficiency and SNR.

The effects of coil deformation must also be considered. As $RF_\text{Stretch}$ conforms to the anatomy, the serpentine conductor path deforms and the spacing between adjacent turns changes. These geometric changes may reduce transmit efficiency and $B_1^+$ homogeneity relative to a rigid solenoid. Their overall effect nevertheless appears modest. As shown in Table \ref{tab:eff_RFs}, the transmit-efficiency advantage of $RF_\text{Stretch}$ over $RF_\text{C-Litz}$ under phantom loading is 40 versus $37\,\upmu\mathrm{T}/\sqrt{\mathrm{W}}$ in NextMRI and 56 versus $50\,\upmu\mathrm{T}/\sqrt{\mathrm{W}}$ in Physio I. \emph{In vivo}, the corresponding values are 31 versus $30\,\upmu\mathrm{T}/\sqrt{\mathrm{W}}$ and 48 versus $43\,\upmu\mathrm{T}/\sqrt{\mathrm{W}}$, respectively. Thus, the transmit-efficiency advantage is retained under anatomical loading, although it is reduced in NextMRI. These findings indicate that the increased filling factor provided by the stretchable implementation compensates for the effects of coil deformation under the conditions evaluated.

\subsection{Imaging SNR}
The phantom imaging results presented in Figure \ref{fig:img_phantom} and Table \ref{tab:img_phantom} are consistent with the $Q$-factor and transmit-efficiency measurements discussed above. In both scanners, $RF_\text{Stretch}$ provided the highest SNR among the evaluated coils, indicating that the increased filling factor enabled by its close conformity to the phantom can compensate for the electrical penalties associated with the stretchable implementation. The advantage was more pronounced in Physio I than in NextMRI. Relative to the rigid coil with the highest SNR in each scanner, $RF_\text{C-Litz}$ in NextMRI and $RF_\text{B-Litz}$ in Physio I, $RF_\text{Stretch}$ increased the global SNR by approximately 8\,\% and 20\,\%, respectively. The corresponding improvements in regions A and B were 15\,\% and 22\,\% in NextMRI and 14\,\% and 42\,\% in Physio I. The controlled phantom geometry enabled reproducible positioning and close contact between the stretchable coil and the sample, which may have contributed to these differences.

The \emph{in-vivo} results were influenced by the additional effects of anatomical geometry, coil deformation, and positioning variability. In NextMRI (Figure \ref{fig:img_knee}a and Table \ref{tab:img_knee}), $RF_\text{C-Litz}$ provided slightly higher SNR than $RF_\text{Stretch}$. The global SNR of the stretchable coil was approximately 5\,\% lower, with slightly smaller differences in regions A and B. This result is consistent with the small reduction in its transmit-efficiency advantage under knee loading and may also reflect changes in $B_1^+$ homogeneity caused by deformation of the coil around the knee. Nevertheless, $RF_\text{Stretch}$ substantially outperformed $RF_\text{A}$, the larger rigid coil used when the smaller coil cannot accommodate the patient comfortably. Relative to $RF_\text{A}$, the stretchable coil increased global SNR by approximately 18\,\%, with improvements of 21\,\% and 23\,\% in regions A and B, respectively.

In Physio I (Figure \ref{fig:img_knee}b and Table \ref{tab:img_knee}), $RF_\text{Stretch}$ provided the highest global and regional SNR. Relative to $RF_\text{B-Litz}$, the rigid coil with the highest SNR, the stretchable coil increased global SNR by approximately 19\,\%, with improvements of 23\,\% in region A and 6\,\% in region B. The larger improvement in region A may reflect the greater local extension of the stretchable coil and the resulting increase in filling factor. The practical advantage was more pronounced relative to $RF_\text{A}$, the large-diameter rigid coil used when knee flexion or limb dimensions prevent use of the smaller coil. In this comparison, $RF_\text{Stretch}$ increased global SNR by approximately 80\,\%, with improvements of 89\,\% and 99\,\% in regions A and B, respectively.

The \emph{in-vivo} SNR differences should be interpreted cautiously. The manual placement of the ROIs and small differences in knee positioning among acquisitions introduce greater variability than in the phantom experiments. Furthermore, the \emph{in-vivo} evaluation was performed in a single volunteer and was not designed to establish population-level performance. Within these limitations, $RF_\text{Stretch}$ provided SNR comparable to or higher than that of the rigid coil with the highest SNR in each scanner and consistently outperformed the larger rigid coil used when patient positioning or anatomy precludes use of a smaller coil.

\subsection{$B_1$ homogeneity}
An additional consideration is the effect of the stretchable implementation on the transmit RF field distribution. Unlike a conventional rigid solenoid, the wearable coil undergoes mechanical deformation as it conforms to the anatomy. During stretching, the serpentine conductor path progressively straightens, the spacing between adjacent turns changes, and the overall geometry deviates from that of the nominal solenoid. These changes may reduce the homogeneity of the transmitted $B_1^+$ field.

At the same time, anatomical conformity reduces the separation between the coil and the anatomy, thereby increasing the filling factor and electromagnetic coupling. The results obtained in this study suggest that these advantages compensate for potential penalties associated with deformation, as $RF_\text{Stretch}$ provided SNR comparable to or higher than that of the rigid coil with the highest SNR in each scanner. However, SNR depends jointly on receive sensitivity, sample loading, and the spatial distributions of the transmit and receive fields. The present experiments therefore do not isolate the contribution of $B_1^+$ homogeneity to the measured SNR.

\subsection{Practical considerations}
Beyond its imaging performance, the proposed wearable coil offers several practical advantages over conventional rigid coils. It can be placed around the patient's leg before the extremity is positioned inside the scanner, in a manner similar to a garment (Figure \ref{fig:fabrication}f). This simplifies coil placement and avoids the need to pass the extremity through a confined, rigid structure within the scanner (Figure \ref{fig:fabrication}g).

This advantage is particularly relevant for patients with pain or limited joint mobility. In our previous clinical studies using low-field MRI systems \cite{FernandezGarcia2026LowFieldKnee,FernandezGarcia2025ESMRMB}, most patients could not comfortably flex the knee sufficiently to position the leg within the smaller rigid coils with a diameter of 15\,cm. Although these coils provided the highest SNR among the rigid coils evaluated in the present study, larger rigid coils were often required to accommodate patient positioning, with a corresponding reduction in SNR. By conforming to the anatomy, $RF_\text{Stretch}$ reduces the knee and ankle flexion required for coil placement. It provided SNR comparable to that of the rigid coil with the highest SNR in NextMRI and higher SNR in Physio I, while clearly outperforming the larger rigid coil used when positioning within a smaller coil is not feasible. The wearable design therefore combines anatomical adaptability with high SNR and may facilitate imaging in patients with restricted mobility.

Compared to our initial stretchable-coil prototypes \cite{Conejero2025RFcoilsISMRM,Conejero2025ESMRMB}, the present design provides greater mechanical robustness while retaining the flexibility required for anatomical conformity. Its principal practical limitation is the fabrication process: the conductor must be stitched manually onto the textile, making fabrication considerably slower than that of conventional rigid coils. A more efficient production method is therefore needed to make manufacturing feasible within a reasonable time. Future clinical translation will also require certified medical-grade textile materials, which were beyond the scope of this proof-of-concept study.

\subsection{Post-processing}
The post-processing results in Figures \ref{fig:processed_next} and \ref{fig:processed_physio} show that denoising and distortion correction improved the visual appearance of all images but did not eliminate the differences associated with the original acquisition SNR. The relative visual performance of the RF coils remained similar after post-processing: images acquired with $RF_\text{Stretch}$ and the rigid coil with the highest SNR retained greater anatomical detail and contrast than those acquired with $RF_\text{A}$. This observation suggests that the benefits of greater coil sensitivity are preserved after post-processing, although the post-processed images were not compared quantitatively.

No clear artifacts specific to the stretchable coil were identified after post-processing. However, a small contrast variation was visible in the axial STIR images acquired with NextMRI (Figure \ref{fig:processed_next}d), with a similar pattern in the Physio I STIR images (Figure \ref{fig:processed_physio}c). This variation was not quantified and cannot be attributed to a specific mechanism based on the present data. One possible contribution is spatial variation in the transmitted $B_1^+$ field caused by deformation of the coil as it conforms to the anatomy. Differences in anatomy and stretching conditions could therefore produce small variations in image contrast.

These observations emphasize that RF-coil performance remains important for maximizing final image quality, even when denoising, distortion correction, and other image-enhancement methods are applied.

\subsection{Limitations and future work}
The \emph{in-vivo} evaluation was performed in a single healthy volunteer. Although comparable acquisition settings were used for all coils within each scanner, the study does not permit statistical analysis or assessment of intersubject variability. Future studies should include larger cohorts of volunteers and patients encompassing different knee sizes, mobility restrictions, and clinical conditions.

The results obtained with NextMRI and Physio I should not be interpreted as a direct comparison between scanners. The systems operate at different frequencies and use different hardware configurations and imaging protocols. The cross-scanner analysis was intended to determine whether the relative performance trends among RF coils were reproduced in two distinct systems, rather than to compare their absolute performance.

%The present study did not include direct characterization of the spatial distribution of the transmitted $B_1^+$ field. Deformation of the stretchable coil may alter this distribution through changes in the overall coil geometry and the spacing between adjacent turns. Direct $B_1^+$ mapping \cite{borreguero2026qualitative,Borreguero2026ESMRMB} and electromagnetic simulations at different degrees of stretching would provide a more complete characterization of these effects.

A quantitative characterization of the transmit field distribution remains an important direction for future work. Direct $B_1^+$ mapping \cite{borreguero2026qualitative,Borreguero2026ESMRMB} or electromagnetic simulations at different degrees of stretching could clarify the effects of conductor deformation, turn spacing, and coil geometry on field homogeneity. Such simulations are technically challenging because explicitly representing the numerous insulated and transposed strands of the Litz-wire conductor, together with their deformation, would require a highly detailed electromagnetic model. Appropriate simplified or homogenized representations of the conductor should therefore also be investigated.

The comfort and positioning advantages of the wearable coil were assessed only qualitatively. Future studies should evaluate these outcomes systematically in patients and clinical personnel using comfort questionnaires and measurements of positioning time. Further work should also address scalable and reproducible fabrication methods, mechanical durability under repeated stretching, and the integration of certified medical-grade textile materials.

 %%%%%%%%%%%% Conclusions %%%%%%%%%%%%

\section{Conclusions}
We developed and evaluated a wearable, stretchable RF coil for knee imaging in two low-field MRI systems operating at different Larmor frequencies. Its performance reflects two complementary factors: the frequency-dependent electrical properties of the Litz-wire conductor and the increased filling factor enabled by anatomical conformity. The electrical advantage of the selected Litz wire diminished with increasing frequency and became a modest disadvantage above approximately 3.1\,MHz. Nevertheless, the conformity of the stretchable design compensated for, and in most experiments outweighed, this electrical penalty. Consequently, $RF_\text{Stretch}$ provided SNR comparable to or higher than that of the rigid coil with the highest SNR in each scanner.

The principal practical advantage of the wearable design arises when pain, restricted mobility, or lower-limb dimensions prevent comfortable positioning within a small-diameter rigid coil, which is frequently the case in practice. Under these conditions, a larger rigid coil is typically required, with a corresponding reduction in SNR. By conforming directly to the anatomy, $RF_\text{Stretch}$ reduces positioning constraints while providing substantially higher SNR than the larger rigid coil. These findings establish the feasibility of wearable, stretchable RF coils for low-field MRI and support their further evaluation for musculoskeletal imaging, particularly when anatomical conformity and patient positioning are important.

\section*{Acknowledgements}

This work was supported by Instituto de Salud Carlos III through the project ``Development of advanced MR techniques for the rapid diagnostic of prostate cancer'' (PMPTA23/00018), co-funded by the European Union. This study has also been funded by the European Union through the project ``NEXTMRI: Truly portable MRI for extremity and brain imaging anywhere \& everywhere'' (HORIZON-EIC-2023-TRANSITION-01/10113640) and through the project INNVA1/2025/48, ``Resonancia magnética preventiva de bajo campo para la detección temprana de cáncer''. JC acknowledges funding from the Spanish Ministry of Science, Innovation and Universities through a Formación del Profesorado Universitario (FPU) grant (FPU23/01559).

\section*{Author Contributions}
See Table \ref{tab:author_contributions}.

\begin{table}[t]
\centering
\caption{Author contributions. An ``x'' indicates participation in the corresponding task.}
\label{tab:author_contributions}
\resizebox{\textwidth}{!}{%
\begin{tabular}{l*{13}{c}}
\hline
\textbf{Task} &
\textbf{JC} &
\textbf{MFG} &
\textbf{JB} &
\textbf{TGN} &
\textbf{PGC} &
\textbf{LVC} &
\textbf{EGC} &
\textbf{RB} &
\textbf{EP} &
\textbf{PM} &
\textbf{FG} &
\textbf{JA} &
\textbf{JMA} \\
\hline
Coil design \& fabrication & x & & & & & & & x & x & x & & & x \\
Scanner preparation       & x & x & x & & & & & & & & & & \\
Data acquisition          & x & x & x & x & & x & x & & & & & & \\
Data processing           & x & x & x & x & x & & & & & & & & \\
Project conception        & x & & & & & & & & & & & x & x \\
Project management        & & & & & & & & & & & x & x & x \\
Figure production         & x & & & & & & & & & & & & x \\
Paper writing             & x & & & & & & & & & & & x & x \\
Paper revision            & x & x & x & x & x & x & x & x & x & x & x & x & x \\
\hline
\end{tabular}%
}
\end{table}

\section*{Conflict of Interest}
TGN consults for PhysioMRI Tech. FG, JMA, and JA are co-founders of PhysioMRI Tech.

\section*{Ethics Statement}
The study was reviewed by the Ethics Committee of the Spanish National Research Council under research agreement number 276/2025.

\section*{Data Availability Statement}
The data that support the findings of this study are available from the corresponding author upon reasonable request.

\bibliographystyle{ieeetr}
\bibliography{Bibliography}

\end{document}